\documentclass[preprintnumbers,twocolumn,amsmath,amssymb]{revtex4}
\usepackage{dcolumn}
\usepackage{bm}
\usepackage[dvips]{color}
\usepackage{graphicx}

\newcommand{\beqs}{\begin{subequations}}
\newcommand{\eeqs}{\end{subequations}}
\newcommand{\be}{\begin{equation}}
\newcommand{\ee}{\end{equation}}
\newcommand{\ba}{\begin{array}}
\newcommand{\ea}{\end{array}}
\newcommand{\bea}{\begin{eqnarray}}
\newcommand{\eea}{\end{eqnarray}}
\evensidemargin=\oddsidemargin

\begin{document}

\title{The like-sign dimuon charge asymmetry in SUSY models}

\author{J.K.Parry}

 \affiliation{
Kavli Institute for Theoretical Physics China,\\
Institute of Theoretical Physics,\\
Chinese Academy of Science, Beijing 100190, China
}

 \begin{abstract}
We study the new physics (NP) implications of the recently reported
$3.2\sigma$ Standard Model (SM) deviation in the like-sign  
dimuon asymmetry at the Tevatron.
Assuming that new physics only enters the $B_s$ mixing amplitude
we explore the implications for generic new physics, general 
supersymmetric (SUSY) models and also SUSY SU(5). 
In the case of SUSY SU(5) we exploit 
the GUT scale relationship between slepton and squark soft masses
to predict rates for lepton flavour violation (LFV). 
The predicted rates for $\tau\to\mu\gamma$ are found to be 
detectable at future Super-B factories.
 \\[2mm]
PACS:  12.10.-g; 12.60.Jv; 13.35.Dx; 14.40.Nd; 13.20.He \hfill
 \end{abstract}

 \maketitle

\section{Introduction}

The Standard Model (SM), with the phase of the 
Cabibbo-Kobayashi-Maskawa (CKM) 
matrix as the sole source of CP violation, 
predicts a very small violation of CP in $B$ meson mixing.
While observations of $B_d$ mixing conform to this paradigm,
in the $B_s$ system there have begun to emerge hints of 
physics beyond the SM.
The first measurements of $B_s\!-\!\bar{B}_s$ mixing and 
its associated CP phase
have been observed by the Tevatron's $\rm D\emptyset$ 
\cite{hepex0603029,hepex0702030,:2008fj} and CDF
\cite{hepex0609040,Aaltonen:2007he} collaborations. 
Although the mass difference $\Delta M_s$ shows little deviation from what is 
expected in the SM, the CP phase $\phi_s$ observed in 
$B_s\to\psi\phi$ decays is found to deviate from the SM by almost $3\sigma$.

More recently,
the $\rm D\emptyset$ collaboration reported a $3.2\sigma$
SM deviation in the like-sign dimuon charge asymmetry \cite{Abazov:2010hv},
\bea
a^b_{sl}=\frac{N_b^{++}-N_b^{--}}{N_b^{++}+N_b^{--}}
=-(9.57\pm 2.51\pm 1.46)\times 10^{-3}
\eea
where $N_b^{\pm\pm}$ is the number of semileptonic events 
$b\bar{b}\to \mu^{\pm}\mu^{\pm} X$, and the 
SM prediction is given as $a_{sl}^{b,SM}=(-2.3^{+0.5}_{-0.6})\times 10^{-4}$
\cite{SM_asym}.
If we assume that there is negligible CP violation in 
the tree-level decay amplitude then this measurement can be interpreted
as further evidence for large CP violation in $B_s$ mixing.

An earlier result by the CDF collaboration found \cite{CDF:assl}, 
$a^b_{sl}=(8.0\pm 9.0 \pm 6.8)\times 10^{-3}$, which has larger uncertainty
yet is still compatible with the $\rm D\emptyset$ result.
As the Tevatron produces both $B_d$ and $B_s$ mesons this measurement
is a linear combination of the individual  asymmetries,
\bea
a^b_{sl}=(0.506\pm 0.043)a^d_{sl}+(0.494\pm 0.043)a^s_{sl}\, .
\eea

Using the average 
of the CDF and $\rm D\emptyset$ results for $a^b_{sl}$ 
and assuming that new physics doesn't enter $B_d$ mixing we find,
\bea
a^s_{sl}= -(17.2\pm 5.9)\times 10^{-3}\, .
\eea
The direct measurement by
$\rm D\emptyset$ \cite{Abazov:2009wg},
$a^s_{sl}=-(1.7\pm 9.1)\times 10^{-3}$, has larger uncertainty yet 
still agrees with the value derived above. 
Taking the average of these we arrive at the combined result,
\bea
(a^s_{sl})_{\text{combined}}= -(12.7\pm 5.0)\times 10^{-3}
\label{assl}
\eea
which we shall use in the analysis which follows.

The dimuon asymmetry has already been investigated in the
context of generic new physics \cite{GenericNP} 
and specific new physics models,
for example models with flavour changing neutral Higgs \cite{FCNCHiggs},
flavour changing $Z'$ \cite{Zprime}, or Axigluons \cite{Axigluon}
(see also \cite{otherpapers}). 
In this work we examine the implications of the recent measurement
of a dimuon asymmetry in SUSY models. 
We first investigate the implications of the recent measurement of $a^b_{sl}$
in generic new physics before studying the dominant SUSY contributions in 
the mass insertion approximation (MIA).
The preferred mass insertion parameter region is determined for two
sample SUSY input points.
In the context of SUSY SU(5)
we then exploit the GUT relationship of squark and slepton 
mass insertions to make predictions for 
the rates of $\tau\to\mu\gamma$ in light of this latest
measurement.

\section{Constraints on New Physics from $B_s$ mixing and $a^s_{sl}$}

The $\Delta B=2$ transition between $B_s$ and $\bar{B}_s$ mesons is 
defined as,
\bea
\langle B^0_s | {\mathcal{H}}^{\Delta B=2}_{eff} |\bar{B}_s^0  \rangle
=2M_{B_s}M^s_{12}
\eea
where $M_{B_s}$ is the mass of the $B_s$ meson. We can then define the 
$B_s$ mass eigenstate difference as,
\bea
\Delta M_s\equiv M_H^s-M_L^s=2|M_{12}^s|
\eea
and its associated CP phase,
\bea
\phi_s=arg(-M_{12}^s/\Gamma_{12}^s)
\eea
In the Standard Model $M_{12}^s$ is given by,
\be
M_{12}^{s,{\rm SM}}=\frac{G_F^2 M_W^2}{12 \pi^2}
M_{B_s}\hat{\eta}^B\,f_{B_s}^2 \hat{B}_{B_s}
(V_{ts}^* V_{tb})^2\, S_0(x_t)
\ee
where $G_F$ is Fermi's constant, $M_W$ the mass of the W boson,
$\hat{\eta}^B=0.551$ is a short-distance QCD correction. 
The bag parameter $\hat{B}_{B_s}$ 
and decay constant $f_{B_s}$ are non-perturbative quantities and contain
the majority of the theoretical uncertainty. $V_{ts}$ and $V_{tb}$ are elements
of the CKM matrix, 
and $S_0(x_t\equiv \bar{m}_t^2/M_W^2)$ is the usual Inami-Lim function.

The Standard Model predictions for the other parameters relevant to 
$B_s$ mixing are \cite{SM_asym}
\bea
\Delta\Gamma^{s,SM}&=& (0.096\pm 0.039)\text{ps}^{-1}\\
\phi^{SM}_s&=&(4.2\pm 1.4)\times 10^{-3}
\eea 

Generic NP contributions to 
$B_s$ mixing may be parameterized as,
\bea
\Delta M_s&=&\Delta M_s^{\rm SM}\,|1+R_s| \label{totalM}\\
\phi_s&=&\phi_s^{\rm SM}+\phi_s^{\rm NP}=
\phi_s^{\rm SM}+arg(1+r_s\, e^{i\sigma_s})
\label{totalphi}
\eea
where $R_s\equiv r_s\, e^{i\sigma_s}=M_{12}^{s,{\rm NP}}/M_{12}^{s,{\rm SM}}$ 
denotes the relative size of the 
NP contribution. 

From the definition of eq.~(\ref{totalM}) we have the constraint,
\bea
\rho_s\equiv \frac{\Delta M_s}{\Delta M_s^{\rm SM}}
=\sqrt{1+2r_s\cos\sigma_s + r_s^2}
\label{rho}
\eea
for $r_s$ and $\sigma_s$. 
The UTfit analysis 
\cite{Bona:2008jn}
gives $\rho_s$ at the $95\%$ C.L. to be,
\bea
\rho_s&=&\left[0.776,\,1.162\right]
\label{rhos}
\eea

From eq.~(\ref{totalphi})
the phase associated with NP can also be 
written in terms of $r_s$ and $\sigma_s$,

\beqs
\bea
\sin\phi_s^{\rm NP}&=&\frac{r_s\sin\sigma_s}
{\sqrt{1+2 r_s\cos\sigma_s+r_s^2}},\\
\cos\phi_s^{\rm NP}&=&\frac{1+r_s\cos\sigma_s}
{\sqrt{1+2 r_s\cos\sigma_s+r_s^2}}
\label{phiNP}
\eea
\eeqs
Here \cite{Bona:2008jn}
 gives the $95\%$ C.L. constraint,
\bea
\phi_{s}^{NP}&\hspace{-1mm}=&\hspace{-1mm}
\left[-162,\,-102\right]^{\rm o}
\cup\left[-76,\,-12\right]^{\rm o}
\label{phisdNP}
\eea
with a two-fold ambiguity.

In order to consistently apply the above constraints all input 
parameters are chosen to match those used in the analysis
of the UTfit group \cite{Bona:2008jn,Bona:2007vi} with the 
non-perturbative parameters,
\bea
f_{B_s}\sqrt{\hat{B}_{B_s}}&=&270\pm 30 \,{\rm MeV}\\
f_{B_s}&=&245\pm 25 \, {\rm MeV}
\eea

Finally, the asymmetry $a^s_{sl}$ is defined as,
\bea
a^s_{sl}=\frac{2w_s z_s \sin\phi_s}{w_s^2+z_s^2}
\simeq \frac{|\Gamma_{12}^s|}{|M_{12}^{s,SM}|}
\frac{\sin\phi_s}{|1+r_s e^{i\sigma_s}|}
\eea
where $w_s=2|M^s_{12}|/\Gamma_s$ and $z_s=|\Gamma^s_{12}|/\Gamma_s$.
We shall assume that NP enters into the mixing amplitude only and
take $\Gamma^s_{12}\equiv \Gamma^{s,SM}_{12}$.
In our numerical analysis we use the combined value of $a^s_{sl}$
in eq.~(\ref{assl}) to constrain the NP parameter space.

\begin{figure}[h]
 \includegraphics[width=6truecm,clip=true]{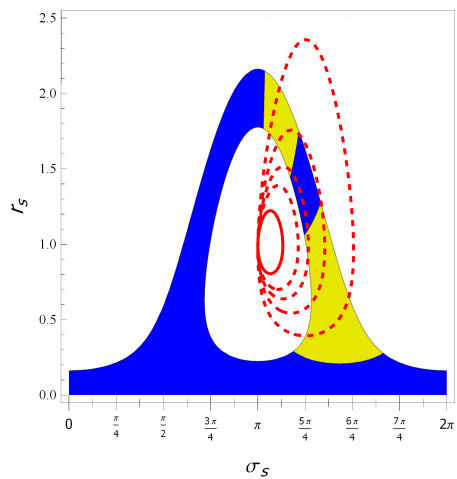}
\vspace{-2mm}
 \caption{The $r_s$-$\sigma_s$ generic new physics parameter space allowed by 
the mass difference $\Delta M_s$ (blue/black) and $\phi_s$ (yellow/light grey) at $95\%$ C.L.. 
The solid red curve corresponds
the central value in eq.~(\ref{assl}), $a^s_{sl}=-12.7 \times 10^{-3}$.
The dashed red curves, from inner to outer curve, correspond to 
$a^s_{sl}=-(7.7,\, 6.0,\, 4.4,\, 2.7)\times 10^{-3}$.}
 \label{fig:r-sigma}
 \end{figure}

Fig.~\ref{fig:r-sigma} shows the $r_s$-$\sigma_s$ NP
parameter space allowed by the constraints
of $\Delta M_s$ and $\phi_s^{NP}$. 
The shaded blue/black region depicts the parameter space 
determined by the $95\%$ C.L. bound from the $\Delta M_s$
measurement, while the two yellow/light grey regions
represent the combined $95\%$ C.L. bounds of 
both $\Delta M_s$ and $\phi_s$. 
Also displayed are red contour lines for the dimuon asymmetry $a^s_{sl}$.
The solid red curve corresponds to the central value
given in eq.~(\ref{assl}) while the $1\sigma$ and $2\sigma$ regions
are shown by the inner most and outer most dashed red curves
respectively. At  the $2\sigma$ level we can see that 
although the two-fold
ambiguity in the CP phase $\phi_s$ isn't resolved by the 
measurement of $a^s_{sl}$, the parameter space is
further restricted. If the CP violation observed in 
the dimuon asymmetry indeed corresponds to NP in 
$B_s$ mixing the tension between $a^s_{sl}$ and $\phi_s$ 
implies that the central value of $a^s_{sl}$
should increase somewhat in the future.

Recently $\rm D\emptyset$ and CDF released new results
based on $6.1 fb^{-1}$ \cite{D0update} and $5.2 fb^{-1}$ \cite{CDFupdate}
of integrated luminosity. These results are still preliminary
and have not yet been combined together. 
As a result we shall not use these 
results in the present analysis. 

In the follow section we study the implications of the 
dimuon asymmetry for the allowed mass insertion parameter space 
of SUSY models and also for the 
predictions of large $\tau\to\mu\gamma$ rates in SUSY SU(5).

\section{The dimuon charge asymmetry in supersymmetric models}

The dominant 
SUSY contribution to $B_s$ mixing comes from the gluino 
contribution which may be written as \cite{hepph0311361},
\bea
R^{\tilde{g}}_s\equiv \frac{M^{s,\tilde{g}}_{12}}{M^{s,SM}_{12}}&=&
a^s_1(m_{\tilde{g}},x)\,
\left[(\delta^d_{RR})^2_{23}+(\delta^d_{LL})^2_{23}\right]
\nonumber\\
&&\hspace{-0.5cm}
+a^s_4(m_{\tilde{g}},x)\,
(\delta_{LL}^d)_{23}(\delta^d_{RR})_{23}+\ldots
\label{Rg}
\eea
where $x$ denotes the ratio of the squared gluino and down-squark masses,
$x=m_{\tilde{g}}^2/m_{\tilde{d}}^2$. The functions $a^s_{1,4}$
can be found in \cite{hepph0311361,Parry:2007fe}.
Here we have ignored terms proportional to $\delta^d_{RL,LR}$
mass insertions as they are expected to be heavily suppressed
due to constraints from $b\to s\gamma$. 

\begin{figure}[h]
 \includegraphics[width=6truecm,clip=true]{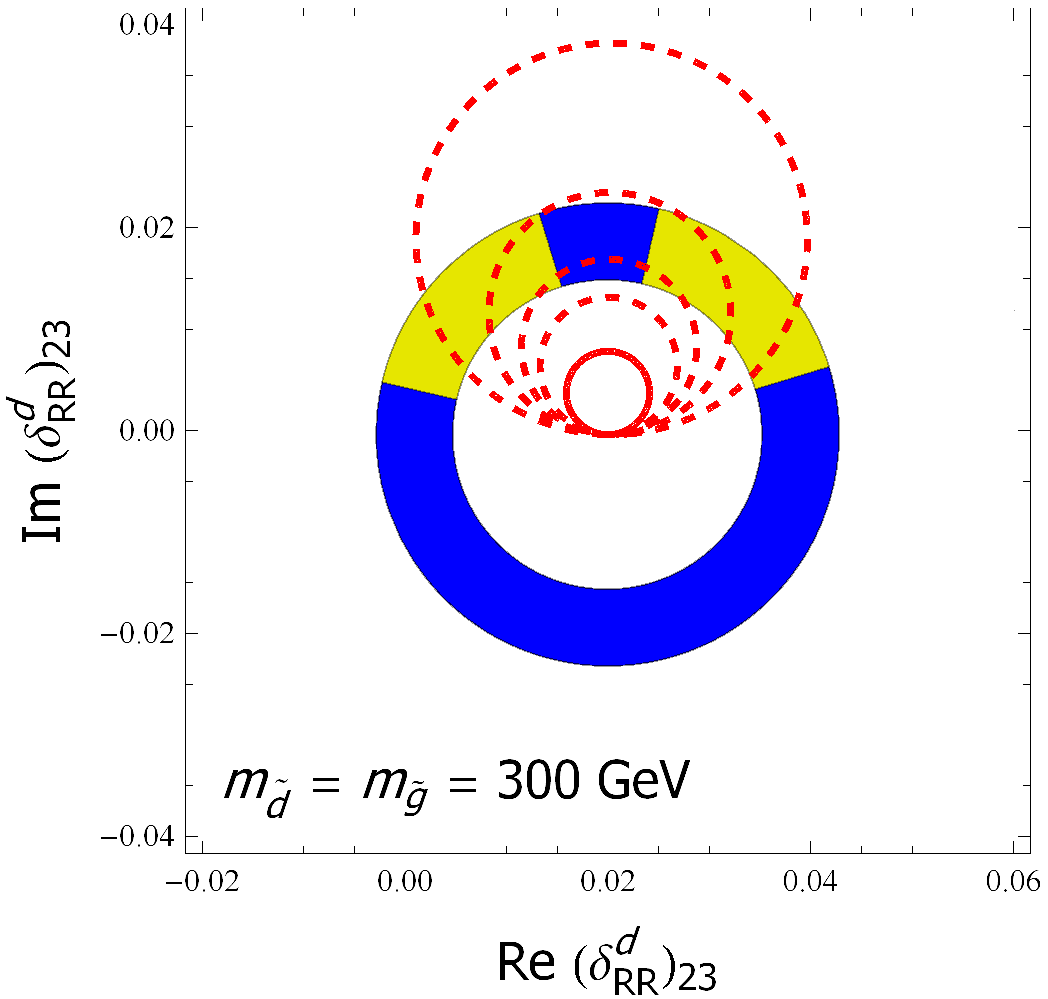}
\\[5mm]
 \includegraphics[width=6truecm,clip=true]{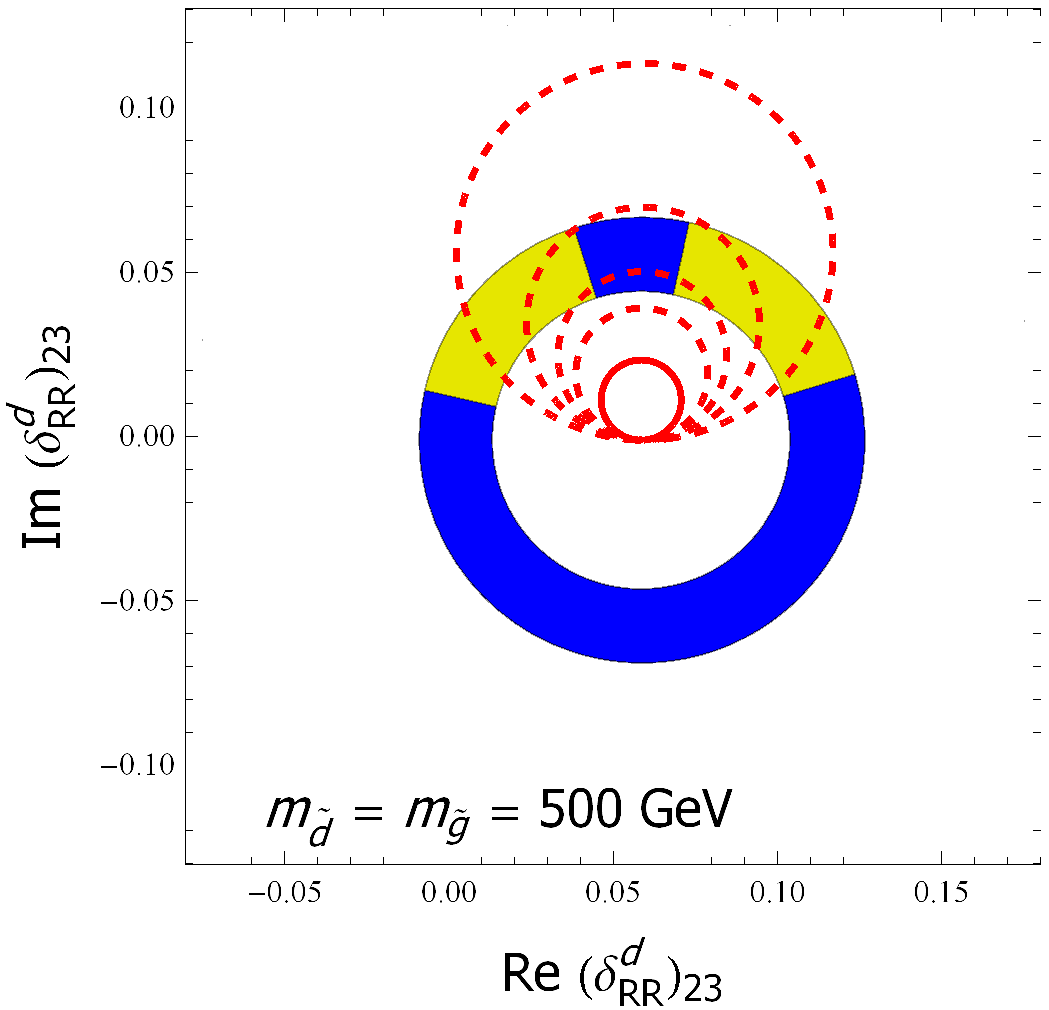}
 \caption{Mass insertion parameter space of $(\delta^d_{RR})_{23}$ 
constrained by the $2\sigma$ measurements of $\Delta M_s$ (blue/black)
and the CP phase $\phi_s$ (yellow/light grey), see eq.~(\ref{phisdNP},\ref{rhos}),
plotted with $m_{\tilde{g}}=m_{\tilde{d}}=300$ GeV (upper) and 
$m_{\tilde{g}}=m_{\tilde{d}}=500$ GeV (lower). 
The red curves show the central value (solid curve),
$1\sigma$ (inner dashed curve) and $2\sigma$ (outer dashed curve) 
constraints from $a^s_{sl}$ as given in eq.~(\ref{assl}).}
 \label{fig:delta}
 \end{figure}

In fig.~\ref{fig:delta} we show the constraints on the 
mass insertion parameter space of $(\delta^{d}_{RR})_{23}$
from the $95\%$ C.L. bounds of the mass difference 
$\Delta M_s$ (blue/black), the CP phase $\phi_s$ (yellow/light grey)
and the dimuon asymmetry $a^s_{sl}$ (red curves).
Again, the solid red curve corresponds to 
the central value of $a^s_{sl}$ from
eq.~(\ref{assl}), while the $1\sigma$ and $2\sigma$ bounds
are shown by the inner most and outer most dashed red curves
respectively. 
At the $2\sigma$ level the new value of $a^s_{sl}$ 
agrees well with both of the regions allowed by the combined
$\Delta M_s$ and $\phi_s$ bounds shown 
by the yellow/light grey shaded areas. This is similar to what we have 
found already in fig.~\ref{fig:r-sigma}.
At the $1\sigma$ level there is a slight tension between 
$a^s_{sl}$ and the CP phase $\phi_s$ which indicates that 
we should either expect the central value of $a^s_{sl}$ to 
increase somewhat in the future, or that new physics also enters into
the width difference, as discussed in \cite{widthdiff}.

In SU(5) the quarks
and leptons are placed into ${\bf 10}=(Q,\, u^c,\, e^c)$ and 
${\bf \bar{5}}=(L,\, d^c)$ representations. 
Due to the symmetry of this simple SUSY GUT
there exists relations amongst the slepton and squark soft 
SUSY breaking masses,
\bea
m_{10}^2=m_{\tilde{Q}}^2=m_{\tilde{U}}^2=m_{\tilde{E}}^2,
\hspace{0.5cm}
m_{5}^2&=m_{\tilde{L}}^2=m_{\tilde{D}}^2
\label{GUT}
\eea
These relations hold for scales at and above the GUT scale. 
Interestingly this implies that left-handed slepton mixing and 
right-handed down squark mixing are related. 
This relation can still be felt
very strongly at the Electroweak scale in the correlation of 
LFV rates and FCNCs.

Due to these GUT scale relations the
$(\delta^d_{RR})_{23}$ contributions to FCNCs and 
$(\delta^l_{LL})_{23}$ contributions to LFV are clearly correlated.
As a result we may explicitly write the rate of $\tau\to \mu\gamma$ 
in the form,
\bea
Br(\tau\to \mu \gamma)\simeq \frac{\alpha^3}{G_F^2}
\frac{m_{\tilde{d}}^4}{M_S^8}|(\delta^d_{RR})_{23}|^2
\tan^2\beta
\eea 
where $m_{\tilde{d}}$ is the average down squark mass and $M_S$ is the 
typical SUSY scale. 

We shall also consider the RGE effects of the CKM mixings in the 
left-handed down squark matrices. At the SUSY scale $M_{SUSY}$
these effects can be approximated as,
\bea
(\delta^d_{LL})_{23}\approx
-\frac{1}{8\pi^2} y_t^2\, V^*_{ts}V_{tb}
\frac{(3 m_0^2+A_0^2)}{m_{\tilde{d}}^2}
\ln\frac{M_{GUT}}{M_{SUSY}}
\eea
Here $m_0$ is the 
universal scalar mass, $A_0$ the universal A-term and $M_{GUT}$ is the
scale of SU(5) unification. 

\begin{figure}[h]
 \includegraphics[width=6.5truecm,clip=true]{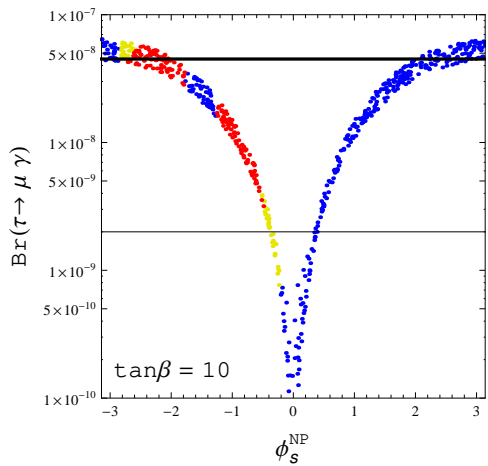}
 \caption{
The variation of the 
branching ratio of $\tau\to\mu\gamma$ with the new physics CP phase
$\phi_s^{NP}$ in SUSY SU(5). The plotted points agree with the $2\sigma$
bounds of $\Delta M_s$ (blue/black), $\phi_s$ (yellow/light grey) 
and $a^s_{sl}$ (red/dark grey).
The bold horizontal line is the experimental bound of
$Br(\tau\to\mu\gamma)< 4.5 \times 10^{-8}$ \cite{hepex0609049}, 
while the narrow line shows the proposed Super-B factory bound
$Br(\tau\to\mu\gamma)< 2 \times 10^{-9}$ \cite{SuperB,SuperKEKB}.
}
 \label{fig:BR}
 \end{figure}

In fig.~\ref{fig:BR} we explore the correlation between 
$B_s$ mixing and the branching ratio for $\tau\to\mu\gamma$
in SUSY SU(5) where we have taken $\tan\beta=10$\footnote{Branching
ratio predictions for $\tau\to e\gamma$ and the ratio 
$Br(\tau\to\mu\gamma)/Br(\tau\to e\gamma)$ have been presented in
\cite{Parry:2007fe}.}. 
The predicted rates for $\tau\to\mu\gamma$ can be scaled 
by $(\tan\beta/10)^2$ for different values of $\tan\beta$.
In \cite{Parry:2007fe} it was found that the 
large CP phase $\phi_s$ restricts the mass insertion parameter
space such that large rates for $\tau\to\mu\gamma$ are 
expected in the case of SUSY SU(5) as shown by the 
yellow/light grey points in fig.~\ref{fig:BR}. From the yellow/light grey 
points we find an approximate lower bound of,
\bea
Br(\tau\to\mu\gamma)\gtrsim 2\times 10^{-9}
\left(\frac{\tan\beta}{10}\right)^2
\eea
The new measurement of
$a^b_{sl}$ further constrains the range of
predictions for $Br(\tau\to\mu\gamma)$. The red/dark grey dots plotted
in fig.~\ref{fig:BR} are those points which agree at the $2\sigma$
confidence level with each of $\Delta M_s$, $\phi_s$ and $a^s_{sl}$.
Including the new measurement we find that the allowed 
points cluster towards the middle of the yellow/light grey band, moving away
from the highest and lowest rates for $\tau\to\mu\gamma$.
Interestingly the majority of the preferred parameter space lies
in the region between the present experimental bound for
$Br(\tau\to\mu\gamma)$ and the reach of the
proposed Super flavour factories \cite{SuperB,SuperKEKB}.

\section{Summary}

In this work we have investigated the impact on 
the new physics parameter space of the 
recent measurement of the like-sign dimuon charge 
asymmetry at the Tevatron.
Assuming that new physics enters only into 
$B_s-\bar{B}_s$ mixing and not into the lifetime difference,
we first saw that this measurement is in reasonably good agreement with 
existing observations of large CP violation in the $B_s$ system.
The viable NP parameter region is further reduced by the latest
Tevatron results although the two-fold ambiguity 
in the CP phase $\phi_s$ remains unresolved.
If new physics present in $B_s$ mixing is indeed responsible for the
dimuon asymmetry we should expect the central value to 
increase in the future.

In the case of generic supersymmetric models and 
supersymmetric SU(5) we have explored the 
mass insertion parameter space and the
correlation between FCNCs and LFV rates in light of the recent 
dimuon asymmetry measurement.
The $(\delta^d_{RR})_{23}$ mass insertion parameter space 
is restricted to two small regions which under the SU(5) GUT relations
predict a $\tau\to\mu\gamma$ branching ratio just below the 
present experimental bound and within the reach of
the proposed Super flavour factories.

\acknowledgments

This work is supported by the National Natural Science Foundation
of China under Grant no. 10875063 and 10875064.


\end{document}